\documentclass[prd,aps,nofootinbib,nobibnotes,superscriptaddress,preprintnumbers,12pt,tightenlines]{revtex4}

\usepackage[dvips]{graphicx}
\usepackage{amsmath,amssymb,mathrsfs}
\usepackage{bm}
\usepackage{times}
\usepackage{epsfig}
\usepackage{verbatim}
\usepackage{bm}

\usepackage{graphics}
\usepackage{graphicx,epsfig,amssymb,amsmath,color, cancel}

\newcommand{\be}{\begin{equation}}
\newcommand{\ee}{\end{equation}}
\newcommand{\ba}{\begin{array}}
\newcommand{\ea}{\end{array}}
\newcommand{\bea}{\begin{eqnarray}}
\newcommand{\eea}{\end{eqnarray}}

\newcommand{\aem}{\alpha_{\rm em}}
\newcommand{\ab}{\alpha_{B}}
\newcommand{\bs}{\boldsymbol}

\begin{document}

\preprint{MCTP-14-08}

\title{New weakly-coupled forces hidden in low-energy QCD}

\author{Sean Tulin}
\affiliation{Michigan Center for Theoretical Physics, University of Michigan, Ann Arbor, MI 48109}

\date{\today}

\begin{abstract}

Is it possible to detect a new weakly-coupled force at the QCD scale that interacts primarily with quarks?  This work investigates experimental signatures of a new MeV -- GeV gauge boson that couples to baryon number, with attention to the 100 MeV -- GeV mass range that is the regime of nonperturbative QCD.   Such a state can be searched for in rare radiative decays of light mesons ($\eta, \eta^\prime, \phi, \omega$) as a $\pi^0 \gamma$ resonance, which is its leading decay mode from $140 - 620$ MeV.  This is a new discovery window for forces beyond the Standard Model that is not covered by existing dark photon searches.

\end{abstract}

\date{\today}

\maketitle

\section{Introduction}

%Fundamental forces in particle physics are described by gauge symmetries.  
%The Standard Model (SM), based on the gauge symmetry $SU(3)_C \times SU(2)_L \times U(1)_Y$, has been extraordinarily consistent with experimental tests of strong and electroweak interactions. Nevertheless, additional gauge symmetries may exist and these forces may be discovered experimentally.  
New gauge symmetries may exist beyond the Standard Model (SM) and these forces may be discovered experimentally.  The stability of dark matter provides one strong motivation for additional forces.  In the absence of a symmetry, it would be puzzling why dark matter particles should not decay on microscopic time scales.  On the other hand, if dark matter is charged under a new gauge symmetry, it is naturally stable, like the electron.

The discovery window for a new gauge boson depends on its mass and how it couples to SM particles.  One model that has been widely considered is a ``dark photon,'' dubbed the $A^\prime$~\cite{Holdom:1985ag} or $U$ boson~\cite{Fayet:2007ua}, that is weakly-coupled to the SM via kinetic mixing and has mass in the MeV -- GeV range.  This scenario has been motivated from many directions, including possible cosmic ray excesses from dark matter annihilation~\cite{Pospelov:2007mp,ArkaniHamed:2008qn,Pospelov:2008jd}, small scale structure anomalies and self-interacting dark matter~\cite{Feng:2009mn, Buckley:2009in,Loeb:2010gj,Tulin:2013teo}, %dark matter direct detection~\cite{Feldstein:2009tr,Chang:2009yt,Fornengo:2011sz,Hooper:2012cw,Kaplinghat:2013yxa}, 
and the muon $g \! - \! 2$ anomaly~\cite{Pospelov:2008zw}.  This, in turn, has inspired a broad program of searching for the $A^\prime$ in high luminosity experiments at the GeV scale, such as $e^+ e^-$ colliders and fixed target experiments~\cite{Reece:2009un,Bjorken:2009mm,Batell:2009yf} (see \cite{Essig:2013lka} for further references).  

Whether or not the anomalies from dark matter and muon $g \! -\! 2$ are taken seriously, it is clear that GeV-scale experiments at the intensity frontier have a unique niche for discovering new forces at intermediate energies that should be fully explored.  However, dark photon searches rely on the {\it leptonic} coupling of this new force, such that the $A^\prime$ may be discovered through its decays to $e^+ e^-$ or $\mu^+ \mu^-$.  Alternative signals have also been proposed based on invisible decays to neutrinos or light dark matter~\cite{Batell:2009di,Artamonov:2009sz,Dharmapalan:2012xp,Davoudiasl:2012ag,Essig:2013vha}.

What if a new force couples predominantly to quarks over leptons?  The simplest model is a gauge boson, dubbed $B$~\cite{Nelson:1989fx}, that couples to baryon number and arises from a new $U(1)_B$ gauge symmetry~\cite{Rajpoot:1989jb,Foot:1989ts,Nelson:1989fx,He:1989mi,Carone:1994aa,Bailey:1994qv,
Carone:1995pu,Aranda:1998fr,FileviezPerez:2010gw,Graesser:2011vj}.  The goal of this work is to explore GeV-scale signatures of the $B$ boson, especially in the range 100 MeV $< m_B <$ 1 GeV that is the domain of low-energy QCD.  

It is well-known that $U(1)_B$ is anomalous given the fermion content of the SM~\cite{Foot:1989ts}.  Consistency of the quantum theory therefore requires introducing new baryonic fermions with electroweak quantum numbers to cancel the $SU(2)_L^2 \times U(1)_B$ and $U(1)_Y^2 \times U(1)_B$ anomalies.  Provided these states are sufficiently heavy, with mass scale $\Lambda \gg m_B$, they may be integrated out of the low-energy effective theory for $B$.  However, $m_B/\Lambda$ cannot be arbitrarily small without the effective theory breaking down~\cite{Preskill:1990fr}.  A ``reasonable'' expectation, assuming that the new fermions acquire masses via a $U(1)_B$-breaking Higgs field, is that $m_B/\Lambda \gtrsim g_B/(4\pi)$~\cite{Williams:2011qb}, where $g_B$ is the $U(1)_B$ gauge coupling.\footnote{This bound is actually much stronger than simply requiring consistency of the effective theory.  With the additional fermions integrated out, anomalies from SM fermions are cancelled by a nonrenomalizable coupling of the $B$ longitudinal mode to the electroweak gauge fields~\cite{Preskill:1990fr}.  This nonrenormalizable coupling required for anomaly cancellation induces radiative corrections to $m_B$ that, according to power-counting arguments~\cite{Preskill:1990fr}, destroy perturbativity of the effective theory unless $m_B \gtrsim g_B g_{\rm SM}^2 \Lambda /(4\pi)^3$, where $g_{\rm SM}$ denotes (generically) the electroweak gauge couplings.  Saturating this bound in a concrete model, however, requires a tiny baryonic charge $q_B \ll 1$ for the $U(1)_B$-breaking Higgs field and $\mathcal{O}(1/q_B) \gg 1$ additional fermions in the ultraviolet theory to cancel anomalies.}  Moreover, requiring $\Lambda \gtrsim 100$ GeV (since no new electroweak fermions have yet been observed in nature) implies that we are targeting very small gauge couplings indeed: $g_B \lesssim 10^{-2} \times (m_B /\mathrm{ 100 \; MeV})$.

It may seem hopeless that a new weakly-coupled baryonic force can be discerned, especially if it preserves the low-energy symmetries of QCD, namely invariance under charge conjugation ($C$), parity ($P$), and $SU(3)$ flavor symmetry.  Nevertheless, there do exist striking signals that can be searched for in high luminosity light meson factories.  

It is important to note that $B$ would not be hidden under the $\rho$ meson since $B \to \pi^+ \pi^-$ violates $G$-parity and is therefore suppressed.  ($G$ is the combined operation of an isospin rotation $e^{i \pi I_2}$ and $C$.)  In fact, the leading decay is $B \to \pi^0 \gamma$ for $m_\pi \lesssim m_B \lesssim 620$ MeV, which leads to distinctive signatures in rare electromagnetic decays of light mesons.  For example, the doubly-radiative decay $\eta \to \pi^0 \gamma \gamma$, which is highly suppressed in the SM with $\mathcal{O}(10^{-4})$ branching ratio, can be mimicked by a new physics decay $\eta \to B \gamma \to \pi^0 \gamma\gamma$.\footnote{Using this decay to constrain a light baryonic force was first pointed out in Ref.~\cite{Nelson:1989fx}. However, it was assumed that $B \to \pi^0 \gamma$ would no longer dominate for $m_B > 2 m_\pi$.}  A Dalitz analysis of the $\pi^0 \gamma$ invariant masses would reveal a peak at $m_B$.  

Searches for long-range forces, from nuclear to macroscopic scales, provide strong limits on a new baryonic force with mass below $m_\pi$~\cite{Barbieri:1975xy,Leeb:1992qf,Adelberger:2003zx,Nesvizhevsky:2007by,Kamyshkov:2008qq}.  On the other hand, the $m_B > {\rm GeV}$ regime has been considered within the context of high energy collider observables, such as heavy quarkonium decay, electroweak observables, and dijet resonance searches~\cite{Carone:1994aa,Bailey:1994qv,Carone:1995pu,Barger:1996kr,Heyssler:1996ji,Aranda:1998fr}.
Studies of $\eta,\eta^\prime,\omega,\phi$ meson decays have a unique capability to cover the gap between these low-mass and high-mass regimes.

In the remainder of this work, these ideas are expanded in further detail.  The $B$ boson model and properties are discussed in Sec.~\ref{sec:Bmodel}, and experimental signatures and constraints are described in Sec.~\ref{sec:constr}.  Our conclusions are summarized in Sec.~\ref{sec:conclusions}.  Calculational details related to $B$ production and decay are given in the Appendix.

%Theoretical models where dark matter carries baryon number have been proposed, which, in addition to stabilizing the dark matter particle, provides an explanation for the coincidence between the dark and baryonic matter abundances in the Universe~\cite{Agashe:2004ci,Farrar:2005zd,Davoudiasl:2010am,Duerr:2013lka}.

\section{$B$ boson model, properties, and meson signatures}
\label{sec:Bmodel}

Before considering a baryonic force, we review the dark photon model.  The $A^\prime$ couples to the SM via kinetic mixing with the usual photon, given by the Lagrangian $\mathscr{L} = - \tfrac{1}{2} \varepsilon F^{\mu\nu} F^\prime_{\mu\nu}$.  Here, $F_{\mu\nu}$ and $F^\prime_{\mu\nu}$ are the photon and dark photon field strengths, respectively, and $\varepsilon$ is the kinetic mixing parameter.  Upon diagonalizing the gauge kinetic terms, the $A^\prime$ acquires couplings to both quarks and leptons proportional to their electric charge.  This coupling is given by $\varepsilon Q_f e$, where $Q_f$ is the electric charge of fermion $f$ in units of the proton charge $e$. 

Turning next to the baryonic force, the interaction Lagrangian is
\be \label{eq:Lint}
\mathscr{L} =  \tfrac{1}{3} g_B  \bar q \gamma^\mu q  B_\mu  \, ,
\ee
where $B_\mu$ is the new gauge field coupling to baryon number.  The gauge coupling $g_B$ is universal for all quarks $q$.  %The normalization factor of $\tfrac{1}{3}$ is a choice of convention.  
We also define a baryonic fine structure constant $\alpha_B \equiv g_B^2/(4\pi)$, analogous to the electromagnetic constant $\aem \equiv e^2/(4\pi) \simeq 1/137$.

Eq.~\eqref{eq:Lint} preserves the low-energy symmetries of QCD.  $C$ and $P$ are conserved, with $B$ being assigned $C\!=\!P\!=\!-$.  Moreover, since the gauge coupling is universal for all flavors, Eq.~\eqref{eq:Lint} preserves $SU(3)$ flavor symmetry acting on $(u,d,s)$.  Of course, $B$ does not transform under the flavor symmetry and is a singlet under isospin.  Thus, $B$ can be assigned the same quantum numbers as the $\omega$ meson: $I^G (J^{PC}) = 0^- (1^{--})$.  

The $\omega$ meson provides a useful guide for how we expect $B$ to decay.  The three leading $\omega$ branching ratios are~\cite{Beringer:1900zz}
\be \label{eq:omegadecays}
{\rm BR}(\omega \to \pi^+ \pi^- \pi^0) \simeq 89\%\,  , \quad {\rm BR}(\omega \to \pi^0 \gamma) \simeq 8\% \, , \quad {\rm BR}(\omega \to \pi^+ \pi^-) \simeq 1.5\% \, .
\ee
The decay $\omega \to \pi^+ \pi^-$ is forbidden by $G$-parity and is suppressed, having to proceed via isospin-violating $\rho$-$\omega$ mixing.  ($\omega \to \pi^0 \pi^0$ is forbidden by $C$.)  
We expect the decay modes of $B$ to be qualitatively similar to \eqref{eq:omegadecays} in the range $m_\pi \lesssim m_B \lesssim {\rm GeV}$.  (At $m_B \approx 1$ GeV, the kaon channel $B \to K \bar K$ opens, and $B$ decays would appear more similar to the $\phi$ meson.)  

In general, $B$ is not completely decoupled from leptons since there should exist kinetic mixing between $B$ and the photon.  If $\varepsilon$ is set to zero at tree-level, say due to a symmetry, one-loop radiative corrections involving heavy quarks generate $\varepsilon \ne 0$~\cite{Carone:1995pu,Aranda:1998fr}.  The typical size of this effect is $\varepsilon \sim e g_B/(4\pi)^2$.  Thus, we consider the more general Lagrangian
\be \label{eq:Lint2}
\mathscr{L}_{\rm int} =  ( \tfrac{1}{3} g_B + \varepsilon Q_q e ) \bar q \gamma^\mu q B_\mu 
-\varepsilon e  \bar \ell \gamma^\mu \ell B_\mu  \, ,
\ee
where $\ell$ is a charged lepton.  Eq.~\eqref{eq:Lint2} includes not only Eq.~\eqref{eq:Lint}, but also dark photon-like couplings proportional to $\varepsilon$.  
The most important effect of $\varepsilon$ is allowing for the decay $B \to e^+ e^-$, which dominates when pion decays are kinematically forbidden.  In this case, $A^\prime$ searches are sensitive to $B$, although $B$ production may be modified compared to $A^\prime$.  

%%%%%%%%%%%%%%%%%%%%%%%%%%%%%%55

\begin{figure}
\includegraphics[scale=0.64]{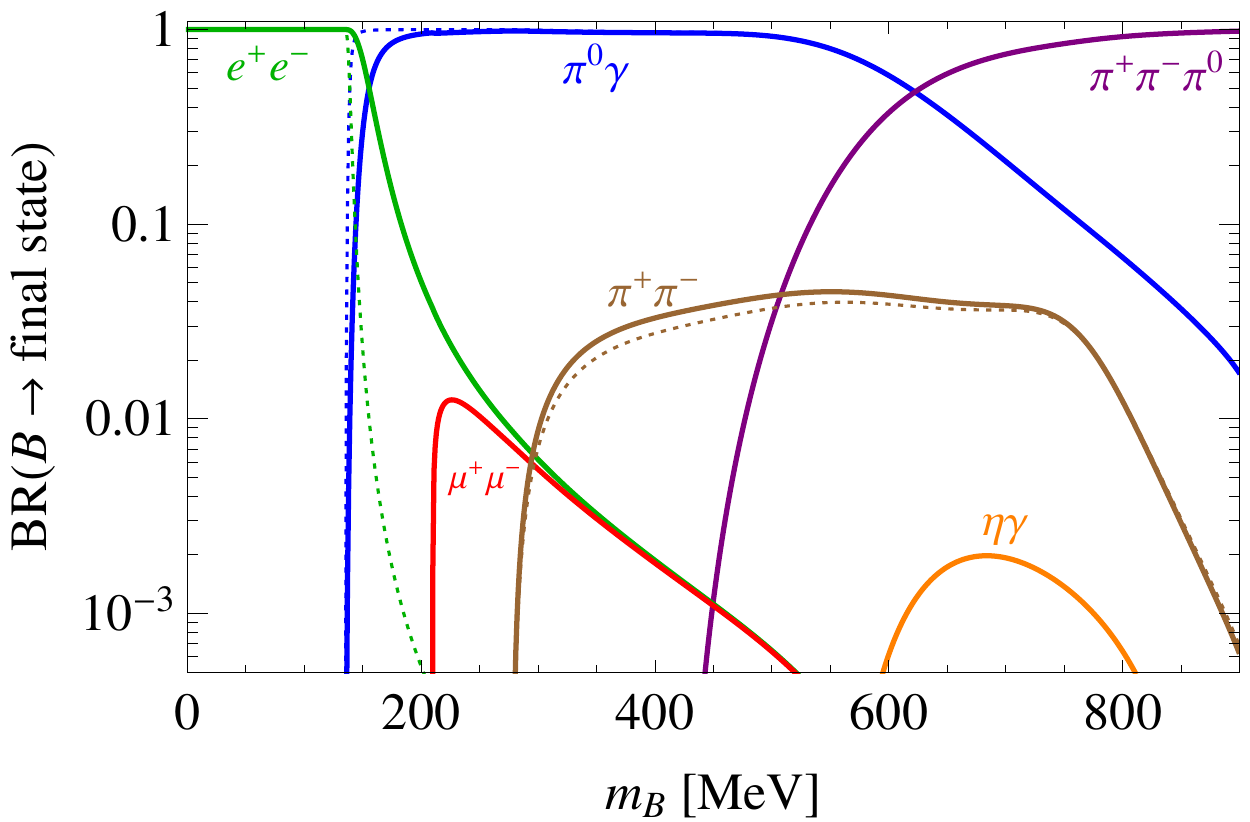}  \includegraphics[scale=0.64]{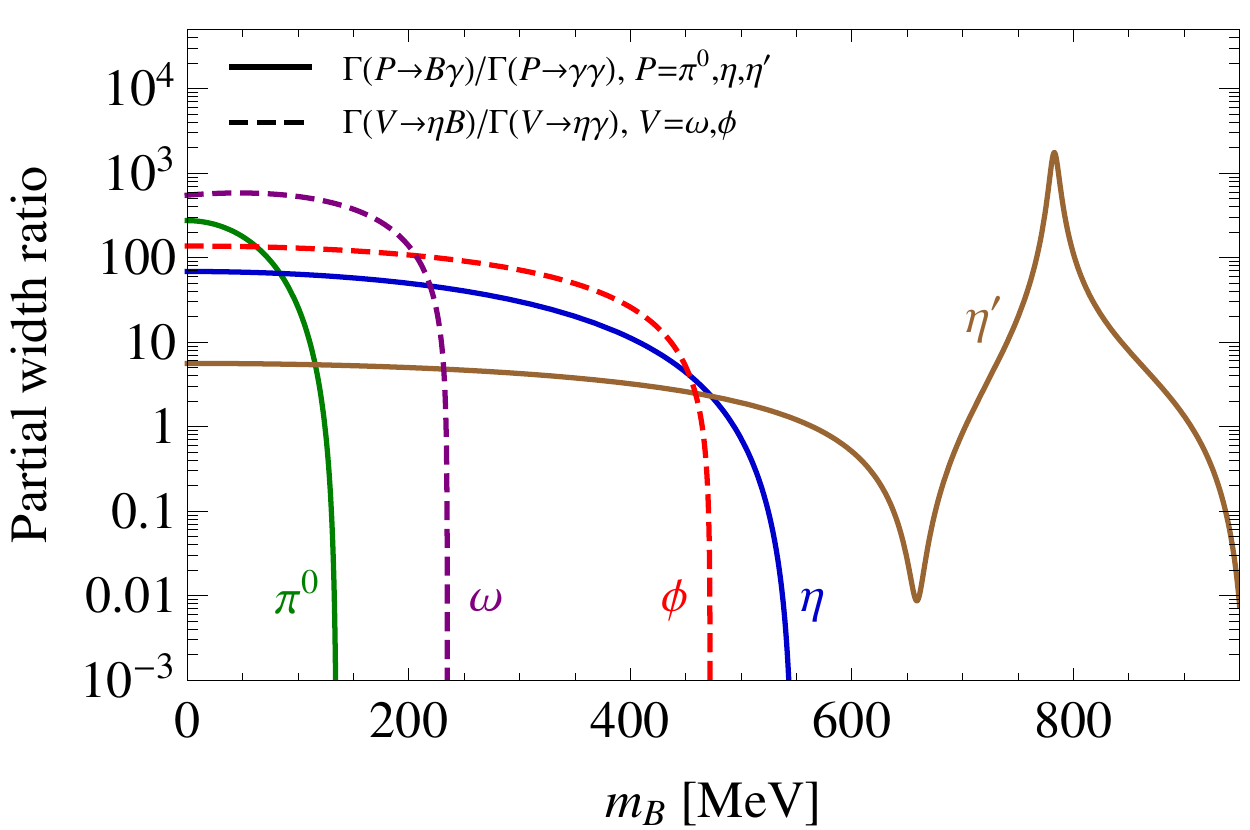}
\caption{\it Left: branching ratios for $B$ decay (independent of $\alpha_B$).  Thick lines have $\varepsilon = e g_B/(4\pi)^2$; thin dotted lines have $\varepsilon = 0.1 \times e g_B/(4\pi)^2$. Right: new physics meson decay widths \eqref{eq:np} relative to the photonic processes \eqref{eq:sm}.  Lines are labeled by the decaying meson.  Solid lines show $\Gamma(\pi^0 \to B \gamma)/\Gamma(\pi^0 \to \gamma \gamma)$, $\Gamma(\eta \to B \gamma)/\Gamma(\eta \to \gamma \gamma)$, and $\Gamma(\eta^\prime \to B \gamma)/\Gamma(\eta^\prime \to \gamma \gamma)$.  Dashed lines show $\Gamma(\phi \to B \eta)/\Gamma(\phi \to \gamma \eta)$ and $\Gamma(\omega \to B \eta)/\Gamma(\omega \to \gamma \eta)$.  Ratios scale proportional to $\alpha_B$ and are shown normalized to $\alpha_B = 1$.   }
\label{fig:Bprop}
\end{figure}

%%%%%%%%%%%%%%%%%%%%%%%%

The partial widths for $B$ decay are computed using vector meson dominance (VMD).  The details of the calculation are given in Appendix~\ref{sec:compute}.  Fig.~\ref{fig:Bprop} (left) shows the resulting branching ratios.  Two values of $\varepsilon$ are considered: a ``natural''-sized value $\varepsilon = e g_B/(4\pi)^2$ that would be induced radiatively (thick lines) and a smaller value $\varepsilon = 0.1 \times e g_B/(4\pi)^2$ that may arise due to a cancellation (thin dotted lines).  The partial widths for $B \to \pi^+ \pi^-$ and $B \to \ell^+ \ell^-$ depend $\varepsilon$, while the $B \to \pi^0 \gamma, \eta \gamma, \pi^+ \pi^- \pi^0$ widths do not.

In terms of branching ratios, our results in Fig.~\ref{fig:Bprop} are summarized as follows:
\begin{itemize}

\item $B \to e^+ e^-$ is the leading decay for $1 \; {\rm MeV} \lesssim m_B \lesssim m_\pi$.  Here, dark photon searches for $A^\prime \to e^+ e^-$ have sensitivity to the $B$ boson as well.  

\item $B \to \pi^0 \gamma$ is the leading decay for $m_\pi \lesssim m_B \lesssim 620 \; {\rm MeV}$.  This is a new channel, not covered in dark photon searches, that may be discovered as a $\pi^0\gamma$ resonance.

\item $B \to \pi^+ \pi^- \pi^0$ is the leading decay for $620 \; {\rm MeV} \lesssim m_B \lesssim 1\; {\rm GeV}$.  Three-pion resonances have been searched for in studies involving $\omega$ mesons.  In fact, for $m_B \sim m_\omega$, the $B$ boson would be a nearly identical twin of the $\omega$.

\item $B \to \pi^+ \pi^-$ is suppressed, with a branching ratio less than $5\%$.  The decay is dominated by $\rho$-$\omega$ mixing with little sensitivity to $\varepsilon$.  In comparison, the dark photon has a sizable $A^\prime \to \pi^+ \pi^-$ branching fraction above threshold.  This stems from the fact that the $A^\prime$ has a different coupling to $u$ and $d$ quarks, due to their different electric charge, thereby violating isospin like the usual photon.

\item $B \to \mu^+ \mu^-$ has a suppressed branching ratio proportional to $\varepsilon^2$. Our result for ${\rm BR}(B \to \mu^+ \mu^-)$ for $\varepsilon = 0.1 \times e g_B/(4\pi)^2$ is not shown in Fig.~\ref{fig:Bprop} and lies below the solid $\mu^+ \mu^-$ line by a factor of 100.

\item $B \to \eta \gamma$ has a small branching ratio.  However, the rate is larger than expected from the $\omega$ meson since $B$ mixes with both $\omega$ and $\phi$.  The ratio $\Gamma(B \to \eta \gamma)/\Gamma(B \to \pi^0 \gamma)$ may perhaps play a role in terms of distinguishing $B$ from $\omega$ in the range $m_B \sim m_\omega$.

\end{itemize}
These results are mostly insensitive to the particular value of $\varepsilon$ chosen,  provided $g_B \gg \varepsilon e$.  The main effect of $\varepsilon$ is to determine the relative branching ratios for $B \to e^+ e^-$ and $B \to \pi^0 \gamma$ near the transition region $m_B \sim m_\pi$.

Next, we turn to how $B$ bosons are produced in light meson decays.  We consider two types of decays, $P \to B \gamma$ and $V \to P B$, where $P$ is a pseudoscalar meson (e.g., $\pi^0, \eta, \eta^\prime$) and $V$ is a vector meson (e.g., $\omega$, $\phi$).  The specific channels of most interest are
\be \label{eq:np}
\pi^0 \to B \gamma \, , \quad \eta \to B \gamma \, , \quad \eta^\prime \to B \gamma \, , \quad 
\omega \to \eta B \, , \quad \phi \to \eta B \, .
\ee
These processes are similar to electromagnetic processes $P \to \gamma\gamma$ and $V \to P \gamma$, namely
\be \label{eq:sm}
\pi^0 \to \gamma \gamma \, , \quad \eta \to \gamma \gamma \, , \quad \eta^\prime \to \gamma \gamma \, , \quad 
\omega \to \eta \gamma \, , \quad \phi \to \eta \gamma \, .
\ee
For $\ab \sim \aem$, the new physics channels \eqref{eq:np} would be comparable to the SM ones \eqref{eq:sm}.  However, relating \eqref{eq:np} and \eqref{eq:sm} is not a simple matter of replacing $\aem \to \ab$ because the photon and $B$ interactions transform differently under $SU(3)$ flavor.  For example, $\omega \to \pi^0 \gamma$ is allowed by isospin, but $\omega \to \pi^0 B$ is not.  In Appendix~\ref{sec:compute}, we compute the partial widths for these processes using VMD.

Fig.~\ref{fig:Bprop} (right) shows the rates for producing $B$ bosons via \eqref{eq:np}, relative to the photonic processes \eqref{eq:sm}.  The solid lines show the ratio $\Gamma(P \to B \gamma)/\Gamma(P \to \gamma\gamma)$ for $P = \pi^0, \eta, \eta^\prime$, while the dashed lines show $\Gamma(V \to \eta B)/\Gamma(V \to \eta \gamma)$ for $V = \omega,\phi$.  The decay widths to $B$ scale proportional to $\alpha_B$ and are shown here normalized to $\alpha_B = 1$.  In other words, the vertical axis shows the value of $\alpha_B^{-1}$ where the new physics decays to $B$ bosons \eqref{eq:np} are equal to the corresponding photonic decays \eqref{eq:sm}.  

It is worth noting the peculiar form for $\Gamma(\eta^\prime \to B \gamma)/\Gamma(\eta^\prime \to \gamma\gamma)$ in Fig.~\ref{fig:Bprop}.  The peak at $m_B \approx m_\omega$ is due to the form factor $F_\omega(s)$ being on resonance, while the valley at $m_B \approx 660$ MeV is due to a cancellation between the $F_\omega$ and $F_\phi$ terms in Eq.~\eqref{eq:etapBg}.  Whether such a cancellation survives beyond the VMD scheme followed here is a useful question for further study since the $\eta^\prime$ has a unique potential to access $m_B$ between $m_\eta$ and $m_{\eta^\prime}$.

%%%%%%
\begin{table}[t]
\begin{tabular}{c|c|c|c|c} 
\hline\hline
\qquad\qquad  Decay $\rightarrow$  & $B \to e^+ e^-$  & $B \to \pi^0 \gamma$  & $B \to \pi^+ \pi^- \pi^0$ & $B \to \eta \gamma$ \\
Production $\downarrow$  \qquad \qquad &\; $m_B \sim 1 - 140$ MeV\;&\; $140 - 620$ MeV \;&\; $620- 1000$ MeV \; &  \\
\hline
$\pi^0 \to B \gamma$ & $\pi^0 \to e^+e^- \gamma$ & -- & -- & -- \\
$\eta \to B \gamma$ & $\eta \to e^+e^- \gamma$ & $\eta \to \pi^0 \gamma \gamma$ & -- & -- \\
$\eta^\prime \to B \gamma$ & $\eta^\prime \to e^+e^- \gamma$ & $\eta^\prime \to \pi^0 \gamma \gamma$ & $\eta^\prime \to \pi^+ \pi^- \pi^0 \gamma$ & $\eta^\prime \to \eta \gamma\gamma$ \\
$\omega \to \eta B $ & $\omega \to \eta e^+e^- $ & $\omega \to \eta \pi^0 \gamma$ & -- & -- \\
$\phi \to \eta B $ & $\phi \to \eta e^+e^- $ & $\phi \to \eta \pi^0 \gamma$ & -- \\
\hline\hline
\end{tabular}
\caption{Summary of rare light meson decays induced by $B$ gauge boson.}
\label{tab:summary}
\end{table}
%%%%

The experimental signatures are a combination of production and decay.  The different possibilities, depending on $m_B$, are summarized in Table~\ref{tab:summary}.  The $e^+ e^-$, $\pi^0 \gamma$, and $\pi^+ \pi^- \pi^0$ decay channels are the dominant modes over the range for $m_B$ indicated, while $\eta \gamma$ is a highly suppressed channel that would be much more challenging experimentally.

%%%%%%%%%%%%%%%%%%%%%%%%%%%55

\section{Signatures and constraints}
\label{sec:constr}

In this section, we discuss searches for the $B$ boson in light meson decays, listed in Table~\ref{tab:summary}, as well as other relevant constraints.  First, we consider the $\pi^0 \gamma$ decay modes.   
\begin{itemize}

\item $\eta\to  \pi^0 \gamma\gamma$: This decay has long been championed as a testing ground for chiral perturbation theory at high order (see Ref.~\cite{Achasov:2001qm} for an excellent historical account).  With this motivation in mind, there has been an experimental resurgence over the past decade in studying this channel~\cite{Knecht:2004gx,Prakhov:2005vx,DiMicco:2005rv,Prakhov:2008zz,Unverzagt:2009vm,Gan:2009zzd,AmelinoCamelia:2010me,Lalwani:2011zz,JEFpro}.
The most precise result published to date is ${\rm BR}(\eta \to \pi^0 \gamma\gamma) = (2.21 \pm 0.53) \times 10^{-4}$ from a reanalysis of data from CrystalBall@AGS~\cite{Prakhov:2008zz}. Preliminary results also exist from KLOE@DA$\Phi$NE~\cite{DiMicco:2005rv}, CrystalBall@MAMI~\cite{Unverzagt:2009vm}, and WASA@COSY~\cite{Lalwani:2011zz}. Future experiments have also targeted this channel~\cite{Gan:2009zzd,AmelinoCamelia:2010me,JEFpro}.  These studies provide an important limit on the $B$ boson as well.  To be conservative, we require that ${\rm BR}(\eta \to B \gamma \to \pi^0 \gamma\gamma) < 3.27 \times 10^{-4}$ so as not to exceed the Ref.~\cite{Prakhov:2008zz} value at $2 \sigma$.  In principle, a much greater discovery reach may be obtained by searching for a peak at $m_B$ in the Dalitz distribution of the two $\pi^0 \gamma$ invariant masses. 

\item $\eta^\prime \to \pi^0 \gamma\gamma$: An upper limit of ${\rm BR}(\eta^\prime \to \pi^0 \gamma\gamma) < 8 \times 10^{-4}$ ($90\%$ CL) from GAMS-2000~\cite{Alde:1987jt}, from nearly three decades ago, remains the only study of this channel.  It is important to note that this limit was obtained by removing the larger resonant $\omega$ contribution, which is of order ${\rm BR}(\eta^\prime \to \omega \gamma) \times {\rm BR}( \omega \to \pi^0 \gamma) \approx 2 \times 10^{-3}$.  To constrain $B$ bosons, we require that this limit is not exceeded in the $m_B < 650$ MeV range to avoid being near the $\omega$ resonance.

\item $\phi \to  \eta \pi^0 \gamma$: This decay, which is expected to be dominated by $\phi \to a_0(980) \gamma \to \eta \pi^0 \gamma$, has long been targeted for its role in elucidating properties of scalar resonances in QCD~\cite{Achasov:1987ts}.  KLOE has provided the most precise branching ratio measurement to date, ${\rm BR}(\phi \to \eta \pi^0 \gamma) = (7.06 \pm 0.22) \times 10^{-5}$, as well as studying the $\eta \pi^0$ invariant mass spectrum as a probe of the intermediate $a_0(980)$~\cite{Ambrosino:2009py}.  As a constraint on $B$, we conservatively require that ${\rm BR}(\phi \to \eta B \to \eta \pi^0 \gamma) < 7.5 \times 10^{-4}$ so as not to exceed the Ref.~\cite{Ambrosino:2009py} value at $2 \sigma$.  However, with KLOE having a sample of $\sim 1.7 \times 10^4$ signal events~\cite{Ambrosino:2009py}, the sensitivity to $B$ is likely to be much improved by searching for a peak at $m_B$ in the $\pi^0\gamma$ invariant mass spectrum.

\item $\omega  \to \eta \pi^0 \gamma$: An upper limit of ${\rm BR}(\omega \to \eta \pi^0 \gamma) < 3.3 \times 10^{-5}$ (90\% CL) was obtained by CMD-2@VEPP-2M~\cite{Akhmetshin:2003rg}, which lies about two orders of magnitude above theoretical SM predictions (see \cite{Radzhabov:2007wk} and references therein).  For this reason, this channel has seen relatively little attention as a probe of hadronic physics.  However, an improved measurement would be useful in constraining the $B$ boson, albeit within a narrow window $m_\pi \lesssim m_B \lesssim m_\omega \!- \! m_\eta$.

\end{itemize}
Next, we consider the two $\eta^\prime$ decay channels for targeting $m_B > 620$ MeV.  For these, it is not clear whether an invariant mass peak can be used to identify $B$ close to the $\rho$ and $\omega$ resonances.  If this is the case, any improvement beyond the limits quoted here would require subtracting off the QCD background.
\begin{itemize}

\item $\eta^\prime \to  \pi^+ \pi^- \pi^0 \gamma$: This channel has been observed at CLEO through charmonium decays in their search for $\eta^\prime \to \omega \gamma \to \pi^+ \pi^- \pi^0 \gamma$, measuring ${\rm BR}(\eta^\prime \to \omega \gamma) = 2.34 \pm 0.30 \, \%$~\cite{Pedlar:2009aa}.  This analysis selected a window for the $\pi^+ \pi^- \pi^0$ invariant mass between $750-814$ MeV.  To place a limit on $B$ bosons in this range, we assume that $\eta^\prime \to B \gamma \to \pi^+ \pi^- \pi^0 \gamma$ does not exceed $2.94\%$.

\item $\eta^\prime \to \eta \gamma\gamma$: No measurement yet exists for this decay, although it has been noted as a possible target for future study~\cite{Gan:2009zzd}.  A process with a resonant $\rho^0$ would be expected at a level of ${\rm BR}(\eta^\prime \to \rho^0 \gamma) \times {\rm BR}(\rho^0 \to \eta \gamma) \approx 9 \times 10^{-5}$.  For illustrative purposes, we consider the possible reach of a constraint on $\eta^\prime \to B \gamma \to \eta \gamma\gamma$ at the level of $10^{-4}$.

\end{itemize}
Our quantitative results for these channels, relying on our calculation for $\eta^\prime \to B \gamma$ in the pole region $m_B \sim m_\omega$, should be regarded with a degree of skepticism.

There exist two other important classes of constraints relevant for constraining $B$ bosons.  We focus here on those that are most relevant in the MeV -- GeV mass range.
\begin{itemize}

\item Searches for long-range nuclear forces are sensitive to the $B$ boson for $m_B \lesssim m_\pi$.  Low-energy nuclear scattering mediated by a long-range force would exhibit a forward-peaked angular distribution, similar to Rutherford scattering.  From the angular dependence measured in keV n-Pb scattering data, Ref.~\cite{Barbieri:1975xy} obtained a limit $\alpha_B < 3.4 \times 10^{-11} \times (m_B/{\rm MeV})^4$ for $m_B \gtrsim$ MeV (see also~\cite{Leeb:1992qf}).

\item Hadronic decays of $\psi,\Upsilon$ quarkonium states, mediated by an $s$-channel $B$ boson, provide constraints on $\alpha_B$ that are independent of $m_B$ for $\sqrt{s} \gg m_B$~\cite{Carone:1994aa,Aranda:1998fr}.  
The strongest limit for $m_B <$ GeV has been obtained from $\Upsilon(1S)$, giving $\alpha_B < 0.014$~\cite{Aranda:1998fr}.

\end{itemize}

Lastly, we consider $B \to e^+ e^-$ signals, which fall within the scope of $A^\prime$ searches.  In a sense, these constraints are orthogonal to other observables since they probe the leptonic couplings of $B$.  Although such couplings are not likely to be absent, their magnitude is subject to an additional model dependence.  Resonance searches by WASA and KLOE have placed stringent limits on ${\rm BR}(\pi^0 \to A^\prime \gamma \to e^+ e^- \gamma)$ and ${\rm BR}(\phi \to A^\prime \eta \to e^+ e^- \eta)$, respectively, where the $A^\prime$ is assumed to decay promptly on detector time scales~\cite{Adlarson:2013eza,Babusci:2012cr}.  
To constrain leptonic $B$ decays, we impose these constraints to the quantities
\be
{\rm BR}(\pi^0 \to B \gamma) \times {\rm BR}( B \to e^+ e^-) \times f_{\rm eff} \, , \quad
{\rm BR}(\phi \to \eta B) \times {\rm BR}( B \to e^+ e^-) \times f_{\rm eff} \, .
\ee
Here, $f_{\rm eff}$ is an (experiment-dependent) efficiency factor that accounts for signal reduction due to nonprompt $B$ decays.  For simplicity, we approximate $f_{\rm eff} \approx 1 - \exp( - \frac{L}{c\tau} )$, where $c\tau$ is the $B$ decay length (neglecting relativistic $\gamma$-factors) and $L$ is the physical scale within which a decay would be considered prompt.  Although limits we present for these channels should be regarded as approximate, we have taken $L=$ 1 cm to be conservative since the true detector geometry is larger~\cite{Adlarson:2013eza,Babusci:2012cr}.

%%%%%%%%%%%%%%%%%%%%%%%%%%%%%%55

\begin{figure}
\includegraphics[scale=.64]{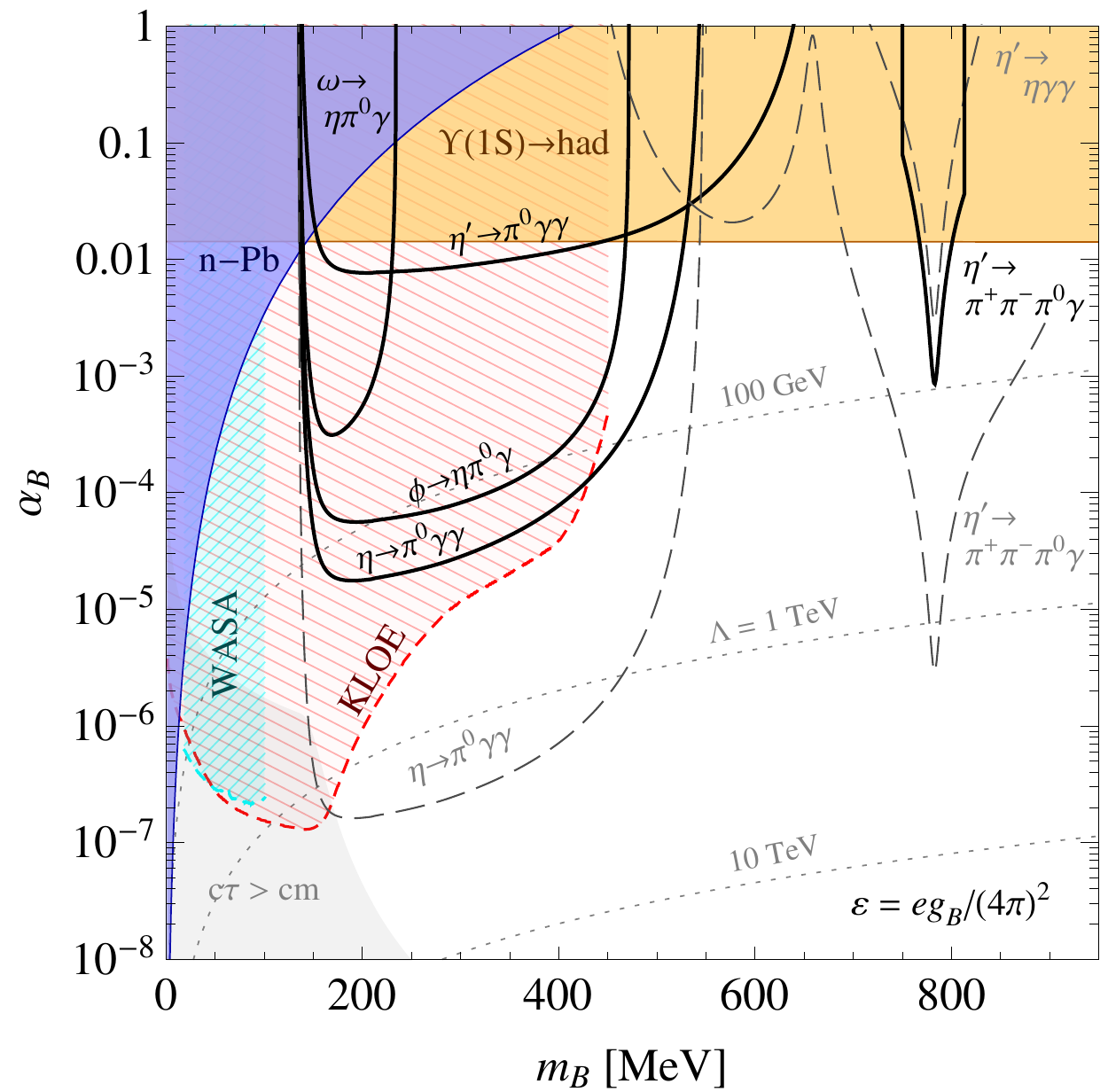} \includegraphics[scale=.64]{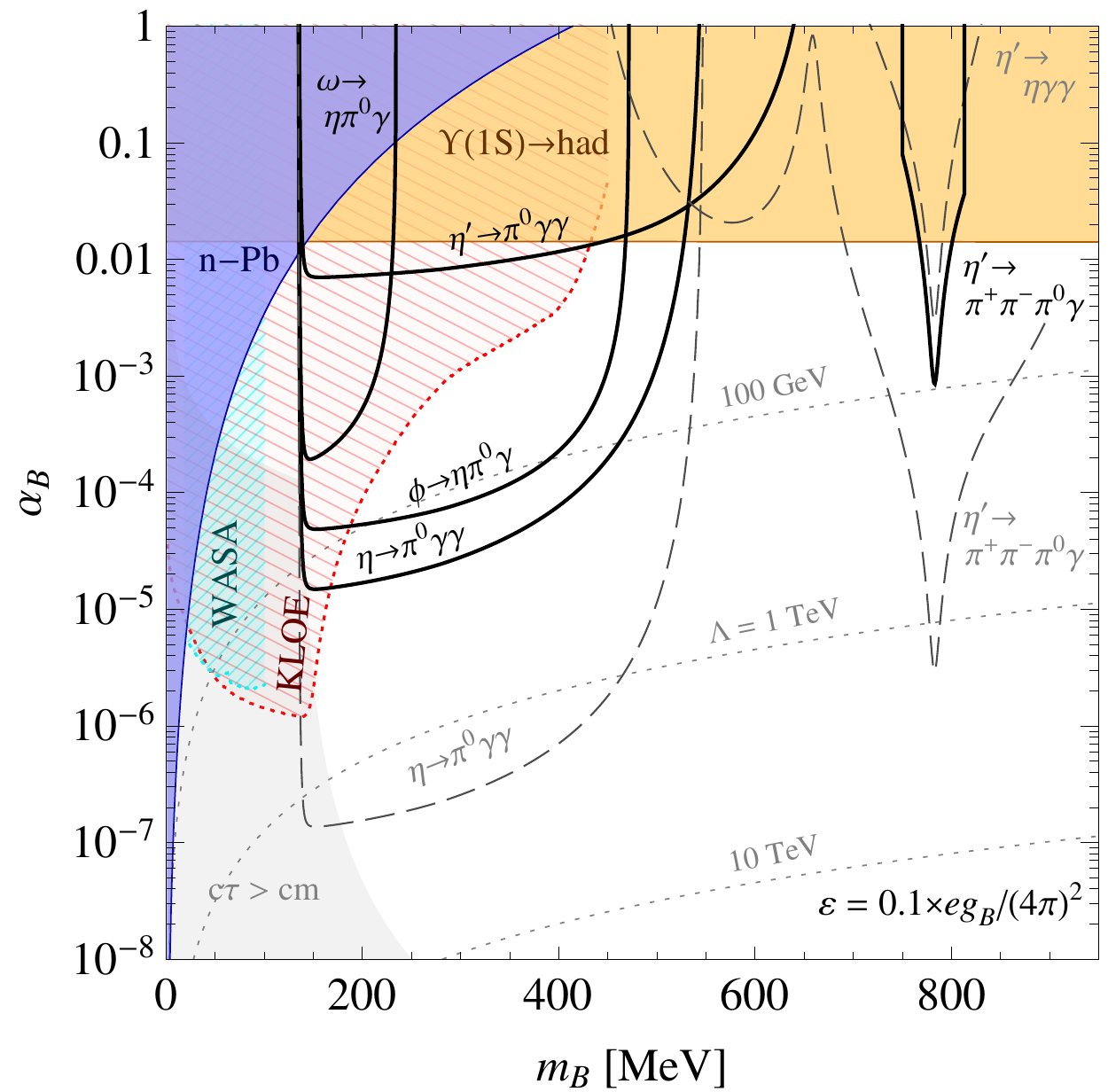}
\caption{\it Limits on baryonic gauge boson coupling $\alpha_B$ and mass $m_B$, for different values of kinetic mixing parameter $\varepsilon$.  Thick black contours are current exclusion limits from radiative light meson decays based on their total rate (assuming the QCD contribution is zero).  Dashed gray contours illustrate the reach of possible future constraints at the level of ${\rm BR}(\eta \!\to\! B \gamma \!\to\! \pi^0 \gamma\gamma)< 3 \times 10^{-6}$~\cite{JEFpro},  ${\rm BR}(\eta^\prime \!\to\! B \gamma \!\to\! \pi^+ \pi^-\pi^0 \gamma)< 10^{-4}$, and ${\rm BR}(\eta^\prime \!\to\! B \gamma \!\to\! \eta \gamma\gamma)< 10^{-4}$.  Shaded regions are exclusion limits from low energy n-Pb scattering and hadronic $\Upsilon(1S)$ decay.  Hatched regions are excluded by $A^\prime$ searches from KLOE~\cite{Babusci:2012cr} and WASA~\cite{Adlarson:2013eza}.  $A^\prime$ limits applied to $B$ are model-dependent, constraining possible leptonic $B$ couplings.  Limits shown here are for $\varepsilon = e g_B/(4\pi)^2$ (left plot) and $0.1 \times e g_B/(4\pi)^2$ (right plot).  Gray shaded regions show where $B$ has a macroscopic decay length $c \tau >$ 1 cm.  Dotted contours denote the upper bound on the mass scale $\Lambda$ for new electroweak fermions needed for anomaly cancellation, assuming $\Lambda \lesssim 4\pi m_B / g_B$.
}
\label{fig:constr}
\end{figure}

%%%%%%%%%%%%%%%%%%%%%%%%

These constraints are shown in Fig.~\ref{fig:constr} in terms of $\alpha_B$ and $m_B$.  The left and right panels correspond to different values of kinetic mixing parameter $\varepsilon$.  The thick black lines, which show how current constraints from radiative light meson decays constrain the $B$ boson, are the new result from this work.  We emphasize that these limits have been applied with respect to the total rate assuming that the QCD contribution is zero.  Substantial improvements could be made by searching for $\pi^0\gamma$ resonances in these processes.

\section{Conclusions}
\label{sec:conclusions}

Light meson decays offer a window into discovering new forces below the GeV scale.  While there exists a broad experimental program of searching for new light weakly-coupled forces, the main focus has been on the dark photon $A^\prime$ and signatures arising from its leptonic couplings.  In this work, we have considered new signatures from a new light force, the $B$ boson, coupled to baryon number.  Although the $B$ boson couples predominantly to quarks, it may be observed in rare radiative decays of $\eta,\eta^\prime,\omega,\phi$ mesons as a $\pi^0\gamma$ resonance.  Such a search may be easily incorporated into the physics programs at existing and future light meson facilities.

Since it is likely for $B$ to couple to leptons at some level, one may wonder: if a new light resonance is observed in $\ell^+ \ell^-$, how can $B$ be distinguished from $A^\prime$?  If the mass is above $m_\pi$, the presence of a $\pi^0\gamma$ resonance signal would be the smoking gun for the $B$ boson.  If the mass is below $m_\pi$, then both $A^\prime$ and $B$ would likely have unit branching ratios to $e^+ e^-$.  In this case, studies of $\omega$ decays can provide a litmus test.  Since $\omega \to \pi^0 A^\prime$ is allowed, while $\omega \to \pi^0 B$ is forbidden by isospin (up to terms suppressed by $\varepsilon$), the absence or presence of an $e^+ e^-$ resonance in $\omega \to e^+ e^- \pi^0$ would favor the $B$ or $A^\prime$, respectively.

We have not discussed possible constraints from beam dump experiments or rare $K$ decays. These topics remain for future study.

\section*{Acknowledgements}

I wish to thank L. Gan, S.~Giovannella, D. Mack, M. Papucci, S. Prakhov, and M. Unverzagt for helpful discussions.  This work benefited greatly from stimulating discussions and hospitality at the Amherst Center for Fundamental Interactions during the ``Hadronic Probes of Fundamental Symmetries'' workshop.  I also acknowledge support from the DOE under contract de-sc0007859 and NASA Astrophysics Theory Grant NNX11AI17G.  

{\it Note added:} While completing this manuscript, Ref.~\cite{Dobrescu:2014fca} appeared, discussing constraints on new fermions related to anomaly cancellation in generalized leptophobic gauge boson models.

%%%%%%%%%%%%%%%%%%%%%5
%
%        APPENDIX
%
%%%%%%%%%%%%%%%%%%%%555

\appendix

\section{Calculation of $B$ decay and production}
\label{sec:compute}

To compute hadronic processes involving $B$, we rely on the hidden local symmetry (HLS) framework for VMD~\cite{Bando:1984ej,Bando:1985rf,Bando:1987br,Fujiwara:1984mp}.  This approach provides a low-energy effective theory describing the pseudoscalar meson nonet $P = (\pi, \eta, \eta^\prime, K, \bar K)$ and the vector meson nonet $V = (\rho, \omega, \phi, K^*,\bar K^*)$ where the latter is treated as a gauge boson of a hidden $U(3)_V$ symmetry.   Since this framework is highly successful in reproducing experimental results for SM decays \eqref{eq:omegadecays} and \eqref{eq:sm}, we are motivated to use it for new physics decays involving $B$.

Let us briefly summarize the HLS-VMD framework.  Following Ref.~\cite{Fujiwara:1984mp}, all processes \eqref{eq:np} and \eqref{eq:sm} can be understood as arising from a single underlying $VVP$ vertex whose coefficient is fixed through the anomaly.   External gauge fields ($\gamma,B$) are included through mixing with $V$.   For the photon, the Feynman rule for $V$-$\gamma$ mixing carries a factor of $e {\rm Tr}[ \bs{Q} \bs{T}_V]$ where $\bs{Q} = {\rm diag}(\tfrac{2}{3}, - \tfrac{1}{3}, - \tfrac{1}{3})$ and $\bs{T}_V$ is the $U(3)$ generator for $V$.  New physics processes \eqref{eq:np} involving $B$ are obtained from the corresponding SM ones \eqref{eq:sm} by replacing
\be
e {\rm Tr}[ \bs{T}_V \bs{Q}] \; \rightarrow \; \tfrac{1}{3} g_B {\rm Tr}[ \bs{T}_V]
\ee
in the matrix element.  For further details, we refer the reader to Ref.~\cite{Fujiwara:1984mp}.

In the remainder of this section, we compute the production and decay rates for processes in Table~\ref{tab:summary}.  The results are summarized in Fig.~\ref{fig:Bprop}.

\subsection{$B$ production from light meson decay}

\begin{figure}
\includegraphics[scale=.9]{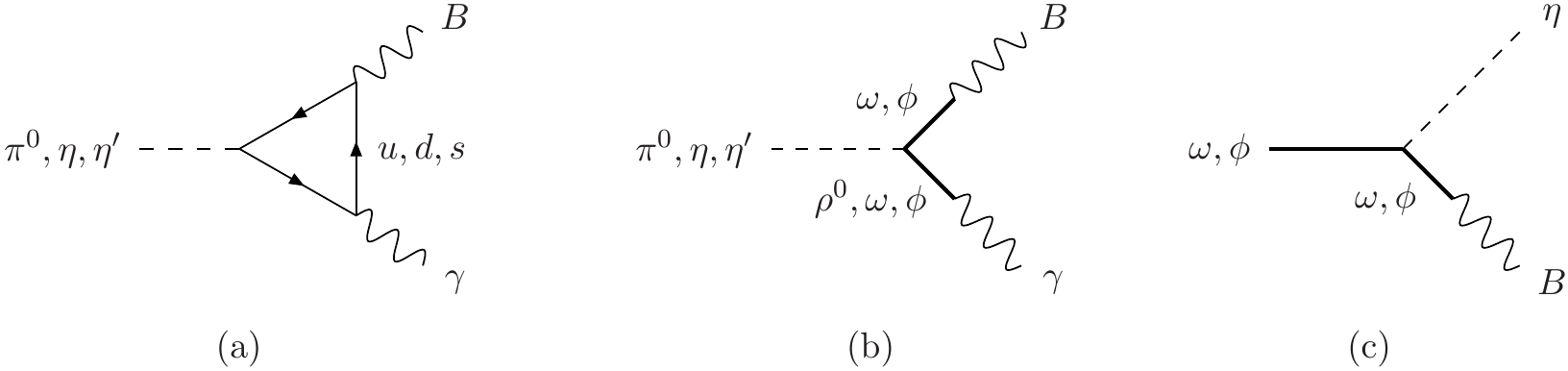} 
\caption{Feynman diagrams for $B$ production via light meson decays.}
\label{fig:feyn2}
\end{figure}

\underline{$P \to B \gamma$:} We consider a pseudoscalar meson $P = \pi^0, \eta, \eta^\prime$ decaying into a $B$ boson with a photon.  Before we turn to the HLS-VMD scheme, it is useful to begin with the more familiar triangle diagram calculation, shown in Fig.~\ref{fig:feyn2}a.  The difference between $P \to B \gamma$ and $P \to \gamma\gamma$ amounts to replacing one factor of $e Q$ by $\tfrac{1}{3} g_B$ in the trace over $u,d,s$ quarks.  The ratio between the two processes is
\be \label{eq:ratio1}
\frac{\Gamma(P \to B \gamma)}{\Gamma(P \to \gamma\gamma)} \approx 2\frac{\ab}{\aem} \left( 1 - \frac{m_B^2}{m_P^2} \right)^3 \left(\frac{1}{3}  \frac{{\rm Tr}(\bs{T}_P \bs{Q} ) }{ {\rm Tr}(\bs{T}_P \bs{Q}^2) } \right)^2   \, ,
\ee
where $\bs{T}_P$ is the generator for $P$.  The explicit form of these generators is 
\be
\bs{T}_{\pi^0} = \tfrac{1}{2} \bs{\lambda}_3 \, , \quad \bs{T}_{\eta} = \tfrac{1}{2} \cos\theta\,  \bs{\lambda}_8 - \tfrac{1}{\sqrt{6}} \sin\theta \, \bs{1} \, , \quad \bs{T}_{\eta} = \tfrac{1}{2} \sin\theta\,  \bs{\lambda}_8 +\tfrac{1}{\sqrt{6}} \cos\theta \, \bs{1} \, ,
\ee
where $\bs{\lambda}_a$ denotes the Gell-Mann matrices and $\bs{1}$ is the identity matrix.  The $\eta$-$\eta^\prime$ mixing angle is $\theta \simeq - 19.5^\circ$, which gives $\sin\theta \simeq -1/3$ and $\cos\theta \simeq 2\sqrt{2}/3$~\cite{Feldmann:1999uf}.   Eq.~\eqref{eq:ratio1} therefore becomes
\be \label{eq:ratio2}
\frac{\Gamma(P \to B \gamma)}{\Gamma(P \to \gamma\gamma)} \approx \frac{\ab}{\aem} \left( 1 - \frac{m_B^2}{m_P^2} \right)^3 \times \left\{ \ba{cc} 2 & P = \pi^0 \\ \tfrac{1}{2} & P = \eta \\ \tfrac{2}{49} & P = \eta^\prime \ea \right. \, .
\ee
$SU(3)$-breaking in the pseudoscalar form factors has been neglected for simplicity.  These effects can be included following Ref.~\cite{Feldmann:1999uf} and amounts to an $\mathcal{O}(1)$ correction to the our results.

The same processes can be computed in the HLS-VMD scheme through the diagram in Fig.~\ref{fig:feyn2}b.  Our results for the $P\to B \gamma$ to $P \to \gamma\gamma$ ratio are
\begin{align}
\frac{\Gamma(\pi^0 \to B \gamma)}{\Gamma(\pi^0 \to \gamma\gamma)} &=
2\frac{\ab}{\aem} \left( 1 - \frac{m_B^2}{m_\pi^2} \right)^3 \left|F_\omega(m_B^2)\right|^2 \label{eq:piBg}\\
\frac{\Gamma(\eta \to B \gamma)}{\Gamma(\eta \to \gamma\gamma)} &= 2 \frac{\ab}{\aem}
\left( 1 - \frac{m_B^2}{m_\eta^2} \right)^3 
\left| \frac{ ( \frac{1}{3} c_\theta  - \tfrac{\sqrt{2}}{3}  s_\theta )  F_\omega(m_B^2)  + (\frac{2}{3} c_\theta + \tfrac{\sqrt{2}}{3}  s_\theta ) F_\phi(m_B^2)  }{  c_\theta  - 2\sqrt{2} s_\theta  } \right|^2 \\
\frac{\Gamma(\eta^\prime \to B \gamma)}{\Gamma(\eta^\prime \to \gamma\gamma)} &= 2 \frac{\ab}{\aem} 
\left( 1 - \frac{m_B^2}{m_{\eta^\prime}^2} \right)^3  
\left| \frac{ ( \frac{1}{3} s_\theta +  \frac{\sqrt{2}}{3}  c_\theta) F_\omega(m_B^2)  + ( \frac{2}{3} s_\theta - \frac{\sqrt{2}}{3}  c_\theta )  F_\phi(m_B^2) }{  s_\theta  + 2\sqrt{2} c_\theta } \right|^2  \label{eq:etapBg}
\end{align}
where $s_\theta \equiv \sin\theta$ and $c_\theta \equiv \cos\theta$.  Form factors $F_{\omega,\phi}(m_B^2)$ arise from the $\omega,\phi$ propagators in Fig.~\ref{fig:feyn2}b and are given by $F_{\omega,\phi}(s) \approx (1 - s/m_{\omega,\phi}^2)^{-1}$ for $s \ll m_{\omega,\phi}^2$.  
However, for the case of $\eta^\prime \to B \gamma$, we require $F_\omega(s)$ in the pole region $s \sim m_\omega^2$.  In this case, we include the $\omega$ width by adopting a Breit-Wigner form $F_\omega(s) \approx 1/(1 - s/m_{\omega,\phi}^2 - i \Gamma_\omega/m_\omega)^{-1}$. The effect from $F_{\omega,\phi}$ is the main qualitative difference from the triangle diagram calculation.  Setting $F_{\omega,\phi} \to 1$, Eqs.~(\ref{eq:piBg}-\ref{eq:etapBg}) reproduce Eqs.~\eqref{eq:ratio1} and \eqref{eq:ratio2} as expected.

%%%%%%%%%%%%%%%%%%%%%%%%%%

\vspace{0.25cm}

\underline{$V \to P B$:} We consider a vector meson $V$ decaying into $B$ along with a pseudoscalar meson $P$.  The most promising decays of this type are $\omega,\phi \to \eta B$ (Fig.~\ref{fig:feyn2}c), which are similar to the photonic decays $\omega,\phi \to \eta \gamma$.  On the other hand, the decays $\omega, \phi \to \pi^0 B$ are less sensitive to $B$ since these channels are suppressed due to isospin violation compared to the isospin-allowed channels $\omega,\phi \to \pi^0 \gamma$. ($\phi \to \pi^0$ transitions are OZI-suppressed as well.) 

The $U(3)$ generators for $V=\rho^0, \omega, \phi$ are, in the ideal mixing limit\footnote{This assumes $\phi \sim s \bar s$ and $\omega \sim \tfrac{1}{\sqrt 2} (u \bar u + d \bar d)$.},
\be
\bs{T}_{\rho^0} = \tfrac{1}{2} \bs{\lambda}_3 \, , \quad \bs{T}_{\omega} = \tfrac{1}{2\sqrt 3}  \bs{\lambda}_8 + \tfrac{1}{3}  \bs{1} \, , \quad \bs{T}_{\phi} = - \tfrac{1}{\sqrt{6}}  \bs{\lambda}_8 +\tfrac{1}{3\sqrt{2}}  \bs{1} \, .
\ee
The form of these generators fixes the relative ratio between $B$-$V$ and $\gamma$-$V$ mixing, given by
\be
\left(\frac{\tfrac{1}{3} g_B {\rm Tr}[ \bs{T}_{V} ] }{e {\rm Tr}[ \bs{T}_{V} \bs{Q} ] }\right)^2 = \frac{\ab}{\aem} \times \left\{ \ba{cc} 4 & \quad V = \omega \\ 1 & \quad V = \phi \\ 0 & \quad V = \rho^0  \ea \right. \, . \label{eq:mixratio}
\ee
From Eq.~\eqref{eq:mixratio}, it is straightforward to compute the relative ratio between partial widths for $\omega,\phi \to \eta B$ compared to the SM decays $\omega,\phi \to \eta \gamma$.  The results are
\bea 
\frac{\Gamma(\omega \to \eta B)}{\Gamma(\omega \to \eta \gamma)} &=& 4 \frac{\ab}{\aem} \frac{\lambda(m_\omega,m_\eta,m_B)^{3/2}}{\lambda(m_\omega,m_\eta,0)^{3/2}} \left| F_\omega(m_B^2) \right|^2 \\
\frac{\Gamma(\phi \to \eta B)}{\Gamma(\phi \to \eta \gamma)} &=& \frac{\ab}{\aem}  \frac{\lambda(m_\phi,m_\eta,m_B)^{3/2}}{\lambda(m_\phi,m_\eta,0)^{3/2}} \left| F_\phi(m_B^2) \right|^2 \, , 
\label{eq:phietaB}
\eea
where $\lambda(m_1, m_2,m_3) = (m_1^2 - (m_2 + m_3)^2)(m_1^2 - (m_2 - m_3)^2)$ is the usual kinematic factor for two-body decay.  The overall numerical factors come from Eq.~\eqref{eq:mixratio} for $V = \omega, \phi$.

Lastly, it is clear from Eq.~\eqref{eq:mixratio} that $B$ does not mix with $\rho^0$.  This provides another demonstration for why $B \to \pi^+ \pi^-$ is forbidden (in the isospin-conserving limit).

\begin{figure}
\includegraphics[scale=1]{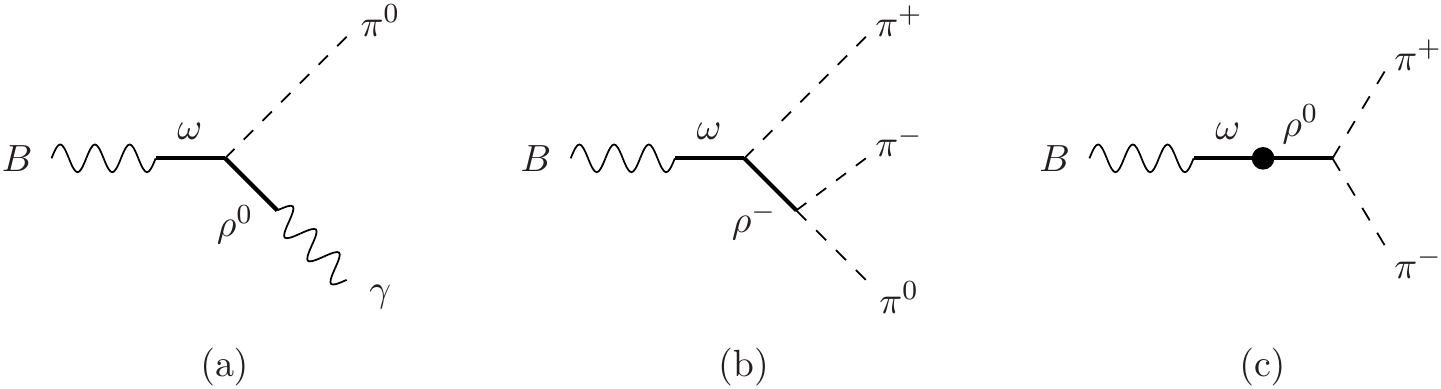} 
\caption{Feynman diagrams for hadronic $B$ decay.}
\label{fig:feyn1}
\end{figure}

%%%%%%%%%%%%%%%%%%%%%%%%%%%%%%%%%%%%%%

\subsection{$B$ decay}
 
\underline{$B \to \pi^0 \gamma$:}  The calculation is identical to that of $\omega \to \pi^0 \gamma$ given in Ref.~\cite{Fujiwara:1984mp}, except for the addition of $B$ mixing with the $\omega$.  The Feynman diagram for this process is shown in Fig.~\ref{fig:feyn1}a.  Neglecting kinetic mixing, the partial width is
\be \label{eq:Bpi0g}
\Gamma(B \to \pi^0 \gamma) = \frac{\alpha_B \aem m_B^3}{96 \pi^3 f_\pi^2} \big( 1 - m_\pi^2/m_B^2 \big)^3 |F_\omega(m_B^2)|^2 \, ,
\ee
where $f_\pi \simeq 93$ MeV is the pion decay constant.   Eq.~\eqref{eq:Bpi0g} is in perfect agreement with an alternative calculation proceeding directly from the usual triangle diagram, shown in Fig.~\ref{fig:feyn2}a, except that $F_\omega(m_B^2)$ is replaced by $F_\omega(0) = 1$.

%%%%%%%%%%%%%%%%%%%%%%%%

\vspace{0.25cm}
\underline{$B \to \pi^+ \pi^- \pi^0$:} Again, we follow the calculation in Ref.~\cite{Fujiwara:1984mp} for the similar process $\omega \to \pi^+ \pi^- \pi^0$.  The relevant Feynman diagram is shown in Fig.~\ref{fig:feyn1}b, plus permutations with an intermediate $\rho^+$ and $\rho^0$.  The end result is
\be \label{eq:B3pi}
\Gamma( B \to \pi^+ \pi^- \pi^0) = \frac{g_{\rho \pi \pi}^4 \alpha_B m_B}{192 \pi^6 f_\pi^2} \mathcal{I}(m_B^2)
|F_\omega(m_B^2)|^2 \, .
\ee
Here, $g_{\rho\pi\pi}$ is the $\rho \pi \pi$ coupling, which is fixed by $g_{\rho \pi\pi}^2/(4\pi) \simeq 3.0$ to give the observed $\rho \to \pi \pi$ decay rate.  The integral over phase space is rather complicated and can be expressed as
\bea
\mathcal{I}(m_B^2) &=& \int \! dE_+ \, dE_- \, 
\big[ |\mathbf{p}_+|^2 |\mathbf{p}_-|^2 - (\mathbf{p}_+ \cdot \mathbf{p}_-)^2 \big] \notag \\
&\;& \quad \times \left( \frac{1}{m_\rho^2 - (p_+ + p_-)^2} + \frac{1}{m_\rho^2 - (p_+ + p_0)^2} 
+ \frac{1}{m_\rho^2 - (p_0 + p_-)^2} \right)^2 \, , \label{eq:Ifunc}
\eea
where $p_\pm = (E_\pm, \mathbf{p}_{\pm} )$ and $p_0 = (E_0, \mathbf{p}_{0} )$ are the $\pi^\pm$ and $\pi^0$ momenta, respectively, in the $B$ rest frame.  All kinematic variables in the integrand are fixed in terms of $E_\pm$, $m_B$, and $m_\pi$.
The phase space integral is restricted to lie in the kinematically allowed domain, which can be expressed as $\int dE_+ dE_- = \int_{m_\pi}^{\epsilon_*} dE_+ \int_{\epsilon_1}^{\epsilon_2} dE_-$, where
\be \label{eq:Ifunclim}
\epsilon_* = \frac{m_B^2 - 3 m_\pi^2}{2 m_B} \, , \quad  \epsilon_{1,2} = \frac{1}{2} \left(m_B - E_+ \pm |\mathbf p_+| \sqrt{ \frac{m_B^2 - 2 E_+ m_B - 3 m_\pi^2}{m_B^2 - 2 E_+ m_B + m_\pi^2} } \right) \, .
\ee
From here, it is straightforward to evaluate Eq.~\eqref{eq:Ifunc} by numerical integration.

%%%%%%%%%%%%%%%%%%%%%%%%

\vspace{0.25cm}
\underline{$B \to \ell^+ \ell^- $:} The leptonic partial width, arising due to kinetic mixing with the photon, is 
\be
\Gamma(B \to \ell^+ \ell^-) = \frac{\aem \varepsilon^2 m_B }{3} \big(1 + 2 m_\ell^2/m_B^2\big) 
\sqrt{1 - 4 m_\ell^2/m_B^2} \, .
\ee

%%%%%%%%%%%%%%%%%%%%%%%%

\vspace{0.25cm}
\underline{$B \to \pi^+ \pi^- $:} The $\pi^+ \pi^-$ decay rate is
\be
\Gamma(B \to \pi^+ \pi^-) = \frac{\aem \varepsilon^2 m_B}{12} (1 - 4 m_\pi^2/m_B^2)^{3/2} |F_\pi(m_B^2)|^2 \, .
\ee
The pion form factor can be expressed as
\be \label{eq:Fpi}
F_{\pi}(s) = F_\rho(s) \left[ 1 + \frac{1+\delta}{3} \frac{\tilde{\Pi}_{\rho\omega}(s)}{s - m_\omega^2 + i m_\omega \Gamma_\omega} \right] \, ,
\ee
where $F_\rho$ is the pion form factor solely due to $\rho$ exchange and $\tilde{\Pi}_{\rho\omega}$ is the additional isospin-violating $\rho$-$\omega$ mixing term~\cite{Gardner:1997ie}.  Taking $\delta=0$ for the moment, Eq.~\eqref{eq:Fpi} is the familiar electromagnetic pion form factor entering $e^+ e^- \to \pi^+ \pi^-$.  Using $e^+ e^- \to \pi^+ \pi^-$ data, Ref.~\cite{Gardner:1997ie} has obtained a phenomenological fit to $F_\rho(s)$ in the time-like region ($s > 0$) and extracted the $\rho$-$\omega$ mixing parameter $\tilde{\Pi}_{\rho\omega}(m_\omega^2) = - 3500 \pm 300 \; {\rm MeV}^2$.  

For $B$ decays, there is an additional term in $F_\pi$, shown in Fig.~\ref{fig:feyn1}c, due to the direct mixing between $B$ and $\omega$.  This contribution is $\delta = 2g_B/(\varepsilon e)$, which only enters $B$ decays and not the electromagnetic process $e^+ e^- \to \pi^+ \pi^-$.  It is important to note that $\delta$ is not a small correction to Eq.~\eqref{eq:Fpi} since $\delta \sim 4\pi/\aem \gg 1$.  In fact, the $\rho$-$\omega$ mixing term is typically the dominant contribution to $B \to \pi^+ \pi^-$. This requires knowing $\tilde{\Pi}_{\rho\omega}(s)$ as a function of $s$, not just at $\sqrt{s} = m_\omega$, and, unfortunately, it is not possible to extract $\tilde{\Pi}_{\rho\omega}(s)$ from $e^+ e^- \to \pi^+ \pi^-$ except at $\sqrt{s} \approx m_\omega$ where $\rho$-$\omega$ mixing is non-negligible.  For lack of a better understanding, we assume here a constant value $\tilde{\Pi}_{\rho\omega}(s) \approx \tilde{\Pi}_{\rho\omega}(m_\omega^2)$, although this assumption is not well-justified.\footnote{It is expected that $\tilde\Pi_{\rho\omega}(s)$ vanishes at least as fast as $s$ for $s \to 0$~\cite{O'Connell:1995wf}.  Therefore, taking $\tilde{\Pi}_{\rho\omega}(s)$ to be constant can be interpreted as a conservative upper bound to $B \to \pi^+ \pi^-$.  If we adopt the ansatz $\tilde\Pi_{\rho\omega}(s) = \tilde\Pi_{\rho\omega}(m_\omega^2) (s/m_\omega^2)$, which vanishes at $s \to 0$ as desired, the branching fraction $B \to \pi^+ \pi^-$ is reduced sizably.}  This limitation precludes our ability to compute $B \to \pi^+ \pi^-$ reliably.

In our numerical calculation, we adopt the ``fit A'' parametrization for $F_\rho(s)$ given by in Ref.~\cite{Gardner:1997ie} obtained by fitting to $e^+ e^- \to \pi^+ \pi^-$ data. However, our results are not strongly sensitive to this choice.  Adopting instead the theoretical calculation of $F_\rho(s)$ from Ref.~\cite{Gounaris:1968mw} or even taking a simple Breit-Wigner form $F_\rho(s) \approx (1 - s/m_\rho^2 - i \Gamma_\rho/m_\rho)^{-1}$ provides only a small correction to our results.

\bibliography{Bforce}

\end{document}